\documentclass{sig-alternate}
\usepackage{amssymb}
\usepackage{leftidx}
\usepackage{ulem}
\usepackage{booktabs}
\usepackage{graphicx}
\usepackage{algorithm} 
\usepackage{algorithmic}
\usepackage{multirow} 
\usepackage{amsmath}
\usepackage{xcolor}
\usepackage{epsfig}
\usepackage{epstopdf}
\usepackage{subfigure}

\begin{document}
%
\conferenceinfo{WOODSTOCK}{ }

\title{ Distributed Computation Particle PHD filter}

%
%
%
%
%

\numberofauthors{4} 
%
\author{
%
%
\alignauthor
Wang Junjie \\
       \affaddr{Harbin Institute of Technology}\\
       \affaddr{P.O. Box 319}\\
       \affaddr{ No.92, West Da-zhi Street, Harbin, China}\\
       \email{jjwanghit@aliyun.com}
\alignauthor
Zhao Lingling\\
       \affaddr{Harbin Institute of Technology}\\
       \affaddr{P.O. Box 319}\\
       \affaddr{ No.92, West Da-zhi Street, Harbin, China}\\
       \email{ zhaolinglinghit@126.com}
\alignauthor  Su Xiaohong\\
      \affaddr{Harbin Institute of Technology}\\
       \affaddr{P.O. Box 319}\\
       \affaddr{ No.92, West Da-zhi Street, Harbin, China}\\
       \email{larst@affiliation.org}
\and  
\alignauthor Ma Peijun\\
       \affaddr{Harbin Institute of Technology}\\
       \affaddr{P.O. Box 319}\\
       \affaddr{ No.92, West Da-zhi Street, Harbin, China}\\
       \email{lleipuner@researchlabs.org}
}

\maketitle
\begin{abstract}

Particle probability hypothesis density filtering has become a promising means for multi-target tracking due to its capability of handling an unknown and time-varying number of targets in non-linear non-Gaussian system. However, its computational complexity grows linearly with the number of measurements and particles assigned to each target, and this can be very time consuming especially when numerous targets and clutter exist in the surveillance region. Addressing this issue, we present a distributed computation particle PHD filter for target tracking. Its framework consists of several local particle PHD filters at each processing element and a central unit. Each processing element takes responsibility for part particles but full measurements and provides local estimates; central unit controls particle exchange between processing elements and specifies a fusion rule to match and fuse the estimates from different local filters. The proposed framework is suitable for parallel implementation and maintains the tracking accuracy. Simulations verify the proposed method can provide comparative accuracy as well as a significant speedup with the standard particle PHD filter.

\end{abstract}



\keywords{PHD filter,SMC,distributed}

\section{Introduction}
 Multi-target tracking(MTT) is  to jointly  estimate the number of targets and position from a  set of uncertain observations.
 The classical approaches such as the nearest neighbor(NN)\cite{singer1971NN},Joint Probabilistic Data Association filter(JPDA)\cite{JPDA1983sonar},  and  Multi-hypothesis tracking(MHT)\cite{blackman2004multiple} are based on the association algorithms.

    Recently,much work has devoted to random finite set(RFS)-based approximations such as the probability hypothesis density(PHD)\cite{maler:phd},\cite{vo:gmphd},cardinalised PHD(CPHD)\cite{mahler:cphd} \cite{vo:cphd} and multiple target multi-Bernoulli(MeMBer) filter\cite{vo:Mfilter}.These methods avoid the data association problem.
 From implementation perspective, a particle PHD filter for nonlinear non-Gaussian MTT problems,was proposed in \cite{vo:smcphd}. And Vo \cite{vo:gmphd} proposed a close-form solution to PHD filter with assumptions on linear Gaussian system. It is called GM-PHD filter.

 The demands of "real-time" MTT have been growing.
 Since the CPHD filter propagates both the intensity of the RFS and the posterior cardinality distribution,its real time characteristic is intrinsically not as good as the PHD filter.The MeMBer filter is more suitable for low clutter environments.And the GM-PHD is constrained to linear Gaussian system. The particle PHD filter is more suitable for nonlinear non-Gaussian MTT problems in dense clutter environment.However,the particle PHD filter computational complexity is very high,Therefore,we are interested in improving the real-time performance of the particle PHD filter.

 Similar to the particle filter,the resampling is chief obstacles to parallel and distribute.  \cite{bolic2005resampling} had proposed the distributed particle resampling algorithm and implement in many WSN applications.In this paper,we proposed a distributed particle PHD filter.
 In order to apply particle PHD filter theory to practice,distributed  algorithm is better than centralized algorithm. Since the distributed algorithm can reduce the execution time by implementing the particle PHD filter using multiple processing elements. We distribute the particles to N PEs.Each PE can run particle PHD filter independently.All individual PEs compute local estimates based over observations in parallel,and transmit their estimate to a CU to obtain global estimate.
 
 Moreover,if the PEs were let to run as independent particle PHD filter,each of them most likely have performance degradation caused by  part of particles.The particles exchange will solve the problem.


 The main contribution of this paper is summarized as follows.
 First,we proposed the distributed particle PHD filter architecture.
 Second,we use the stphd method to extract the state estimation and corresponding measurement label.

  The real time performance is enhanced and the tracking performance is equal to traditional particle PHD filter.

 The remainder of this paper is organized as follows, standard particle PHD filter is given in section 2.Section 3 ,we present our distributed particle PHD filter in detail.Simulation results are given in Section 4.Section 5 is devoted to the conclusions.

\section{ Backgroud}
\subsection{ The PHD FILTER}
 The PHD filter was developed in the framework of Finite Set Statistics(FISST) initially.
 The PHD function $D_{\Xi}$ is the first order moment of the random finite set(RFS) $\Xi$
 and can be  defined by

 \begin{equation}
    D_{\Xi}(x) \equiv E[\delta_{\Xi}(x)] = \int \delta_{X}(x)P_{\Xi}(dX)
 \end{equation}

where $\delta_{\Xi}(x) = \sum_{x \in \Xi} \delta_{x}$ is the random density representation of $\Xi$. $P_{\Xi}$ is the probability distribution of the RFS $\Xi$. The PHD has the properties that, the integral over a measurable subset $S\subseteq E$ $\int_{S} D_{\Xi}(x)\lambda(dx)$ is the expected number of target.In addition,the peaks of the PHD
function give the estimates of the target states.

PHD filter consists of the   \textit{prediction} step and the  \textit{update} step.Assuming the RFS is Poisson, it has been shown that the recursion propagating the PHD $D_{k|k}$ of the multi-target posterior $p_{k|k}$  follows \cite{maler:phd}

\begin{equation}
   D_{k|k} = (\Psi_k \circ \Phi_{k|k-1})D_{k-1|k-1}
\end{equation}

where $\circ$ represents composition of functions,$ \Phi_{k|k-1}$  is the prediction operator and $ \Psi_k$  is the update operator,which are defined as follows:

\begin{equation}
(\Phi_{k|k-1}\alpha)(x) = \gamma_k + \int \phi_{k|k-1}(x,\xi)\alpha(\xi) \lambda(d\xi)
\end{equation}

\begin{equation}
 (\Psi_k \alpha)(x) = \left[ 1 - P_D(x) + \sum_{z \in Z_k} \frac{\psi_{k,z}(x)}{\kappa_k(z)+<\psi_{k,z},\alpha>} \right] {\alpha(x)}
\end{equation}

for any function $\alpha$ on $E_s$,where

 \centerline{$
   \phi_{k|k-1}(x,\xi) = e_{k|k-1}(\xi)f_{k|k-1}(x|\xi)+b_{k|k-1}$}

\centerline{$
   \psi_{k,z}(x) = p_D(x)g_k(z|x)$}

\centerline{$
   \kappa(z)=\lambda_k c_k(z)$}

\centerline{$
   <f,g>=\in f(x)g(x)\lambda(dx)$}

\subsection{Particle PHD Filter}

 As an approximate implementation of PHD filter,the particle PHD filter is composed of three steps:

At time $k>0$,let $L_k$ and $J_k$ denote the number of survival particles and birth particles at time k,repectively\\
1)Prediction step:

For i=1,...,$L_{k-1}$,sample $  \widetilde{x}_k^i \sim q_k(.|x_{k-1}^i,Z_k)$ and compute the predicted weights
\begin{equation}
  \widetilde{w}_{k|k-1}^i=\frac{\phi_{k|k-1}(x_k^i,x_{k-1}^i)}{q_k(x_k^i|x_{k-1}^i,Z_k)}w_{k-1}^i
  \label{Eq:predict1}
\end{equation}

For i=$L_{k-1}+1,...,L_{k-1}+J_k$,sample $ \widetilde{x}_k^i \sim p_k(.|Z_k)$ and compute the weights of new-born particles
\begin{equation}
  \widetilde{w}_{k|k-1}^i=\frac{\gamma_{k}(x_k^i)}{p_k(x_k^iZ_k)}\frac{1}{J_k}
  \label{Eq:predict2}
\end{equation}

2)Update step

For each $z\in Z_k$ compute
\begin{equation}
  C_k(z)=\sum_{j=1}^{L_{k-1}+J_k}{\psi_{k,z} \widetilde{w}_{k|k-1}^j* \widetilde{x}_k^j}
\end{equation}

For i=1,...,$L_{k-1}+J_k$ update weights
\begin{equation}
  \widetilde{w}_k^i=[ \nu(\widetilde{x}_k^i)+ \sum_{z\in Z_k}{\frac{\psi_{k,z}(\widetilde{x}_k^i)}{\kappa_k(z)+C_k(z)}}] \widetilde{w}_{k|k-1}^i
\end{equation}

3)resampling step

Compute the total target number $N_k=\sum_{j=1}^{L_{k-1}+J_k}{\widetilde{w}_k^j}$,resample
$\left\{{\widetilde{x}_k^i,\widetilde{w}_k^i/N_k}_{i=1}^{L_{k-1}+J_k} \right\}$ to get $\left\{{x_k^i,w_k^i/N_k}_{i=1}^{L_k} \right\}$

Just like particle filter,the application of particle PHD filter is limited to its computational complexity which is mainly caused by the resampling and it also caused by the update which need all particles participate.


\subsection{Distributed particle filter}
The method of DRNA(distributed resampling with non-proportional allocation) was initially proposed by Bolic().The idea of the DRNA PF is to divide the whole particles into several groups so that the resampling can be performed independently by group and thus be implemented in parallel.The general DRNA is outlined by:

1) Exchange particles among groups

2) For k=1,...K and i=1,...,N sample $x_t^{k,i} \sim \pi(x_t)$ in parallel in each group

3) For k=1,...K and i=1,...,N compute the weights in each group in parallel
\begin{equation}
 w_t^{*(k,i)}=\frac{w_{t-1}^{(k,i)}p(y_t|x_t^{(k,i)})p(x_t^{(k,i)}|x_{t-1}^{k,i})}{\pi(x_t^{(k,i)})}
\end{equation}

4) Normalize the weights of the particles with the sum of the weights in the group

5) Resample inside the groups

\section{Distribution PHD Filter}
The probability hypothesis density(PHD) filter,which propagates only the first moment instead of the full multi-target posterior,still involves multiple integrals with no closed forms generally(by vol).So BA-NGU VO proposed the particle PHD filter.The particle PHD filter is suitable for problems that nonlinear non-Gaussian dynamics.However,it's high computational is to hold back it's application into real time system.
To speed up the particle PHD filter, we propose a distributed approach that only uses a subset of particles to different computing cores. In other words,we use a subset of particles to different PE and acquire the same accuracy as all the particles were used together ,but avoid unnecessary  communication among the PEs. We entitle this idea DCPFPHD in this paper,can be formalized as follows. The structure chart of DCPFPHD algorithm is shown in Fig.1


\begin{figure}
  \centering
  \includegraphics[width=3in]{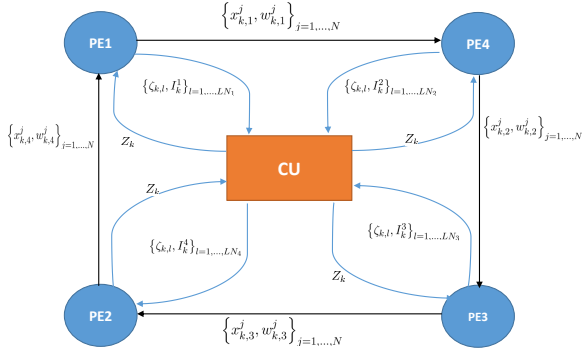}\\
  \caption{the structure of DCPFPHD}
\end{figure}

As the particle PHD filter is similar the particle filter,also involve three basic steps,we can use DRNA in particle PHD filter too.In this paper,we apply the DRNA scheme to particle-PHD filter.However,there is some differences in how calculate the weight and estimate target,the challenge is change the DRNA so we can use it in particle PHD filter(Fig 1).

\subsection {General structure}
The standard particle PHD filter requires all the particles be participated by a (single processor) in resampling step.
 In this paper,Assume we have K PEs and each  can run a separate particle PHD with M particle.The total number of particles is N=MK.
The distributed computation particle PHD filter can be outlines as follows,

Assume we have K groups and each  can run a separate particle PHD with M particle.The total number of particles is N=MK.
The distributed computation particle PHD filter can be outlines as follows,

\begin{algorithm}{DCPFPHD}
    \caption{Depiction on Each Group j}
    \label{alg:DCPFPHD}
    \begin{algorithmic}
         \REQUIRE ~~ $\left\{x_{k-1}^{(i,j)}\right\}_{i=1,...M},Z(k)$ \\
         \textbf{ Local PF-PHD filter}
         \FOR{$i = 1,\dots,M$}
         \STATE $x_k^{(i,j)} \sim q(\cdot|x_{k-1}^{(i,j)},z_k)$.
          \ENDFOR
         \FOR{$i = M+1,\dots,M+J_k$}
         \STATE $x_k^{(i,j)} \sim p(\cdot|z_k)$.
          \ENDFOR
          \FOR{$i=1,\dots,M+J_K$}
          \STATE weight update \\
          $w_k^{(i,j)}=[ \nu(x_k^i)+ \sum_{z\in Z_k}{\frac{\psi_{k,z}(x_k^i)}{\kappa_k(z)+C_k(z)}}]w_{k|k-1}^i$
          \ENDFOR\\
          \textbf{ Local  Estimation}\\
          \STATE estimate  $N_k(j)$ and state(j)\\
          \textbf{ Resample}\\
          \STATE resample the $M+J_k$ particles into  $N_k(j)R_{k}$ particles and set $M=N_k(j)R_k$\\
           \textbf{ Share Particles}\\
           \STATE each group send t particle-weight couples to neighbors

    \end{algorithmic}
\end{algorithm}

      In the following,we describe the algorithm steps in detail.

      t=0:Generate M particles for each group.The  number of  all particles is N.All these particles are shared with same weights $\omega_0=1/M$

      t=t+1

      Step 1:Prediction   At time t-1,we assume the particle set $ x_{t-1}^{(k,m)}$.$w_{t-1}^{k,m} $ is available.For k-th group and m=1,...$L_k$,sample $\left\{ x_t^{(k,m)} \right\}$ from $q_k$. For new-born particles we divide these particles to K group equal then join the groups.

       Step 2:Update In update step,we can calculate the weights among groups.Let $Z_k$ denote the measurements set.For each group,for each  $z in Z_k$,use the likelihood and compute $C_t^ k $ and then update the weights.

       Since we use the STPHD method  to estimate state,the update step should make some change.zhao\cite{zhao2010new} has demonstrated that  the $D_{k|k}$ can be calculated as

\begin{equation} \label{eq:stphd}
   D_{k|k}(x|Z_{1:k}) = \sum_{z \in Z_k} \Delta D_{k|k}(x|z) + \Delta D_{k|k}(x|\phi)
\end{equation}
where $ \Delta D_{k|k}(x|\phi)$ denotes the PHD of the measure undetected.
For each observation $z_p \in Z_k$ ,$p=1,...,M_k$,For each group j,j=1,...,K
\begin{equation}
    C_k(z_{k,p})=\sum_{i=1}^{M+J_k}{\psi_{k,z_{k,p}} \widetilde{w}_{k|k-1}^{(i,j)} \widetilde{x}_k^{(i,j)}}
\end{equation}
\begin{equation}
G_{k}^{i,p,j} = \frac{\psi_{k,z}(\widetilde{x}_{k}^{i,j})}{\kappa(z_{k,p})+C_k(z_{k,p})}
\end{equation}
then calculates the sub-weight of each particles for all observations $z_{k,p}$
\begin{equation}
\widetilde{ \Delta w}_{k}^{i,p,j} = G_{k}^{i,p,j} \widetilde{w}_{k|k-1}^{i,j}
\end{equation}
And the particle sub-weight for the target without measurements obtained is
\begin{equation}
\widetilde{\Delta w}_{k}^{i,p,0} = \nu(\widetilde{x}_k^{i,j})\widetilde{w}_{k|k-1}^{i,j}
\end{equation}
Based on formula   \eqref{eq:stphd} ,the weights can compute through
\begin{equation}
   \widetilde{w}_{k}^{i,j} = \sum_{p=1}^{M_k}\widetilde{ \Delta w}_{k}^{i,p,j} +  \widetilde{\Delta w}_{k}^{i,p,0}
\end{equation}

\paragraph{Step 3:Local estimation}

We use the STPHD method can extract the estimate targets and these targets' sequence number.
For each measurement $z_{k,p}$,$p=1,...,M_k$,compute the sum of $\widetilde{ \Delta w}_{k}^{i,p,j}$ relevant to $z_{k,p}$ in group j.

\begin{equation}
   \Delta W_{k}^{j,p} = \sum_{i=1}^{M+J_{k}} \widetilde{ \Delta w}_{k}^{i,p,j}
\end{equation}

Compute the sum of sub-weight $\Delta W_{k}^{j,0}$ corresponding to targets without observations:
\begin{equation}
   \Delta W_{k}^{j,0} = \sum_{i=1}^{M+J_{k}} \widetilde{ \Delta w}_{k}^{i,0,j}
\end{equation}

Since the weight sum of all the particles equals to the target number,the local target number can be estimated by
$
   LT_{k}^{j} = round(\sum_{i=1}^{i=M}\widetilde{w}_{k}^{i,j})
$
where $round(\sum_{i=1}^{i=M}\widetilde{w}_{k}^{i,j})$ is the nearest integer to $\sum_{i=1}^{i=M}\widetilde{w}_{k}^{i,j}$

Find the $LT_{k}^{j}$ largest sum weight  $\Delta W_{k}^{j,p}$ and the index set $I_{k}^{j} $ relevant to $\Delta W_{k}^{j,p}$ . The local estimated target state can be calculated by
$\zeta_{k,l} = w_{k}^{i,l,j} \widetilde{x}_{k}^{i,j}$ where $l$ in index set ,

\begin{equation}
   w_{k}^{i,l,j} = \frac{\widetilde{ \Delta w}_{k}^{i,l,j}}{ \sum_{i=1}^{M+J_k}\widetilde{ \Delta w}_{k}^{i,p,j}}
\end{equation}

 The groups  can   send data to the CU. When the group get a local estimate like $\zeta_{k,l}$ and transmit the pair $\left\{ \zeta_{k,l},I_{k}^{j} \right\}$ to the CU.Then the CU can combine the local estimate state which depends on the measurement index from the group to construct a global estimate.

\paragraph{Step 4:Global Estimation}

As the CU receives all the local information  $\left\{ \zeta_{k,l},I_{k}^{j} \right\}$.Depend on the  rule  "at most one measurement per target"\cite{streit2013probability}.If the local estimate state's label $\left\{I_{j}^{i}\right\} $ is same that from different groups,then we use their mean value as the global estimate state. And the local estimate state may be from clutter,so we consider only the local estimate states' number which have same label greater than   half of groups' number as valid estimate states.

\paragraph{Step 5:Resampling} The resampling can be carry out locally at the N groups.Normalize the weights of the particles with the sum of weights in the group.
\paragraph{Step 6:Local exchange} The particles in the n-th group will degenerate when its aggregated weight becomes negligible relative to the aggregated weights of the other groups.Then the n-th group hardly contribute to the approximation of the posterior probability distribution.In order to keep the groups are valid,neighboring groups can exchange a portion of particles and weights.We select the L particles from k-th group($L<M/2$) to replace the  L particles from k-1 group in random.

for k=1,...K-1,i=1....,L do:

$\left\{ {x_t^{i,k}} \right\} \rightarrow \left\{ {x_t^{i,{k+1}}} \right\}$

for k=K ,i=1....,L do:

$\left\{  {x_t^{i,K}} \right\} \rightarrow \left\{  {x_t^{i,1}} \right\}$

The processors are connected using an interconnection network.There are many type of network can be used,we select a ring configuration in this paper.For a ring network ,the nth PE,n=1,...,N-1,can send data to the (n+1)th PE.The Nth PE transmits data to PE number 1.

The PEs can also send data to the CU. When the PE get a local estimate like $ddd$ and transmit the pair {} to the CU.Then the CU can combine the local estimate state which depends on the measurement index from the PE to construct a global estimate.

\subsection {Analysis of Time Delay and Computational Complexity}
In the traditional particle PHD filter,all the particles have to be involved by serial. In our methods,since particles are divided into K groups and the group can run a particle PHD filter independently,thus it only  utilise $1/K$ time than before in theory.

\section{SIMULATION RESULTS AND EXPERIMENTAL STUDY}
To evaluate the proposed distributed particle PHD filter,we consider a  two-dimensional scenario with the target can disappear and appear at anytime.Each target moves according to the following model
$      
x_{k+1}=\left[                 
  \begin{array}{cccc}   
     1 & T & 0 & 0\\  
     0 & 1 & 0 & 0\\  
     0 & 0 & 1 & T \\
     0 & 0 & 0 & 1\\
  \end{array}
\right] x_{k} +  \left[                 
  \begin{array}{cc}   
     T^2/2 & 0 \\  
     1 & 0 \\  
     0 &  T^2/2\\
     0 &  1\\
  \end{array}
\right]w_k
$
where  $x_k=[ x_k ,  \dot{x}_{k}  , y_k , \dot{y}_{k} ]$  is target state vector at time kT(k is the time index and T=1 is the sampling period).[$x_k$,$y_k$] is the position,while $[\dot{x}_{k} ,\dot{y}_{k}]$ is the velocity.$w_k=[ w_k^x , w_k^y ]$ is the vector of independent zero-mean Gaussian white noise with standard deviations of
$      
 \left[                 
  \begin{array}{cc}   
     0.025 & 0 \\  
     0 & 4 \\  
  \end{array}
\right]
$
There is just a signal sensor in the scenario and the target-originated measurement are given by\\
              $z_k$=g($x_k$)+$v_k$=$[R,\theta]^T$\\
$
 =\left[
  \begin{array}{c}
    R=\sqrt{(x-s_x)^2+(y-s_y)^2} \\
    \theta = \arctan{(y-s_y)/(x-s_x)}\\
  \end{array}
\right] +\left[                 
  \begin{array}{c}
     v_k^R \\
     v_k^\theta\\
  \end{array}
\right]
$

The measurement variance $v_k^R=5 m,v_k^\theta=0.05 rad$.
Clutter is modeled as a Poisson RFS $ \kappa_k$ with intensity
$
   \kappa_k(z)=\lambda_kV_u(z)
$
The Target can disappear or appear in the scene at any time.The probability of target survival is $e_{k|k-1}=0.9$ and is detected with probability $P_{D,k}=1 $ . Assume the target birth according to the Poisson distribution with the intensity  $N(\cdot,\overline{x},Q)$ where  $N(\cdot,\overline{x},Q)$ denotes a normal density with mean $\overline{x}$
  and covariance Q.

$
  \overline{x} = \left(  \begin{array}{c}
                          0\\
                          3\\
                          0\\
                          -3\\
                         \end{array}
                        \right)
$,
$
 Q =  \left(  \begin{array}{cccc}
                          10 & 0 & 0 & 0\\
                          0  & 1 & 0 & 0 \\
                          0  & 0 & 10 & 0\\
                          0  & 0 & 0 & 1\\

                         \end{array}
                        \right)
$

The surveillance region is $[-\pi/2,\pi/2]\times[0,200]$ rad.m.
Assign 200 particles to an exist or new born target in each group.The number of group is 4.
\subsection{Simulation Results}
\begin{figure}
\centering
\subfigure[  ]{
\label{fig:subfig:a} 
\includegraphics[width = 0.3\textwidth,keepaspectratio]{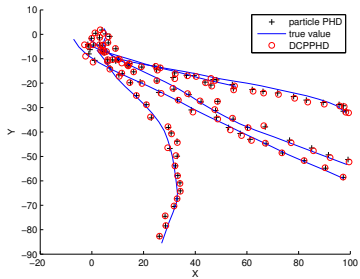}}

\subfigure[   ]{
\label{fig:subfig:b} 
\includegraphics[width = 0.3\textwidth,keepaspectratio]{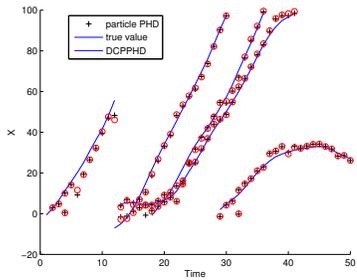}}
 
\label{fig:subfig} 
\subfigure[   ]{
\label{fig:subfig:c} 
\includegraphics[width = 0.3\textwidth,keepaspectratio]{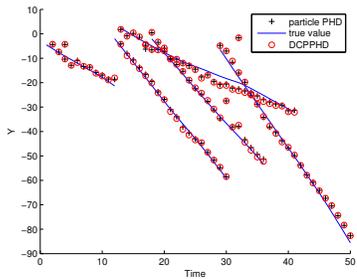}}
\caption{Target trajectories estimated by the DCPPHD filter and the particle PHD filter.(a)Estimated trajectories and true trajectories.(b)Estimated trajectories and true trajectories in x-axis direction.(c)Estimated trajectories and true trajectories in y-axis direction}
\label{fig:subfig} 
\end{figure}

We run 100 independent simulations of the BOT model given by....The overall number of particles was N = 2000 and ,for the DCPFPHD,we divide them into N=4 processors with M=500 particles each.
The true trajectories of five tracks over 50 scans are plotted in Fig \ref{fig:subfig:a}.Fig \ref{fig:subfig:a} also shows the positions of the estimated targets over 50 time steps.
The individual x and y coordinates of the tracks and estimated targets for each time step are shown in Fig \ref{fig:subfig:b} and Fig \ref{fig:subfig:c},respectively.It can be seen that estimated position based on the traditional particle PHD filter which the number of  particles is equals the DPHD filter's particles and the DCPPHD are similar and they are all close to the true tracks.

\begin{figure}
  \centering
  \includegraphics[width=3in]{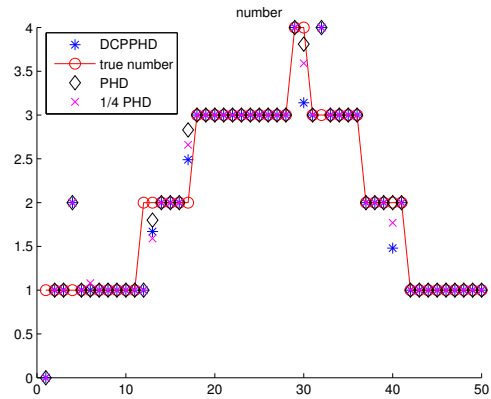}\\
  \caption{ Estimated number of targets
and the true number of targets}\label{10}
\end{figure}

Another vital factor for the performance of the method is the estimated number of targets.The number of true targets and estimated target by our method  ,standard particle PHD filter with same particles and only $1/k$ particles at each scan are given in Fig \ref{10}.It is observed that,under the same simulation conditions and same particles ,the particle PHD filter and DCPPHD filter achieve 

\begin{figure}
\includegraphics[width=3in,height=4.5cm]{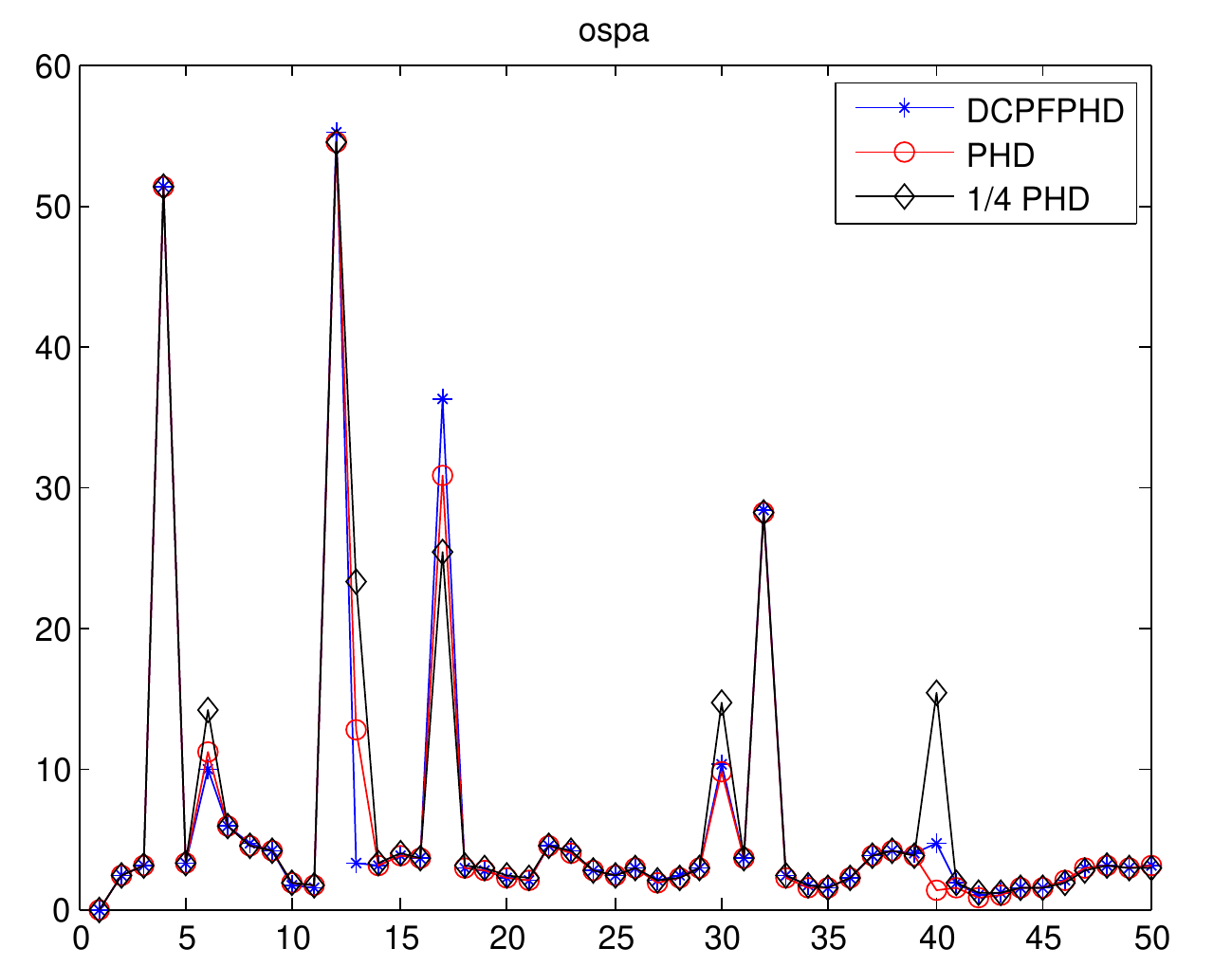}
\caption{the ospa }\label{ospa_10}
\end{figure}

It was proposed in \cite{vo:ospa} to use the Optimal Sub-Pattern Assignment(OSPA) as a multi-target miss-distance metric,and the parameters in it are set as p = 1 and c = 100 inour evaluation. Fig \ref{ospa_10} shows the OSPA distance of the DCPPHD filter and particle PHD filter.
 
The total number of particles for these methods is given in Tabel1.Implemented on a Dell computer using MATLAB,this approximate times used by these approachs are also given in Tabel1.And the Table1 shows the mean of OSPA as well.

\begin{table*}
\centering
\caption{12345}
\begin{tabular}{|l|c|c|c|c|c|} \hline
method & particle number &  times     & mean of OSPA & std of OSPA & r\\ \hline
PHD    & 1 in 1,000      &  17.6540   & 3.0577        &0.0548      & 0  \\ \hline
DCPPHD & 1 in 5          &  8.6331    & 3.0732        &0.0830      & 0   \\ \hline
partPHD& 1 in 40,000     &  4.0935    & 3.1993        &0.2590      & 0    \\ \hline
PHD    & 1 in 1,000      &  72.2214   &  6.4432        & 19.0788     & 10  \\ \hline
DCPPHD & 1 in 5          &  27.1516    & 6.2220        & 5.4630      & 10   \\ \hline
partPHD& 1 in 40,000     &  17.2492    & 7.4752        & 54.4956      & 10    \\ \hline
PHD    & 1 in 1,000      &  117.1504  &  8.0049        & 8.2928      & 20  \\ \hline
DCPPHD & 1 in 5          &  39.4615    &  8.0884        & 22.6434     & 20   \\ \hline
partPHD& 1 in 40,000     &  29.2382    & 9.7143       &  136.3478      & 20    \\ \hline

\end{tabular}
\end{table*}

set simulation results
1.the target position\\
2.the ospa\\
3.the time\\
4.the number\\
\begin{figure}
\includegraphics[width=3in,height=4.5cm]{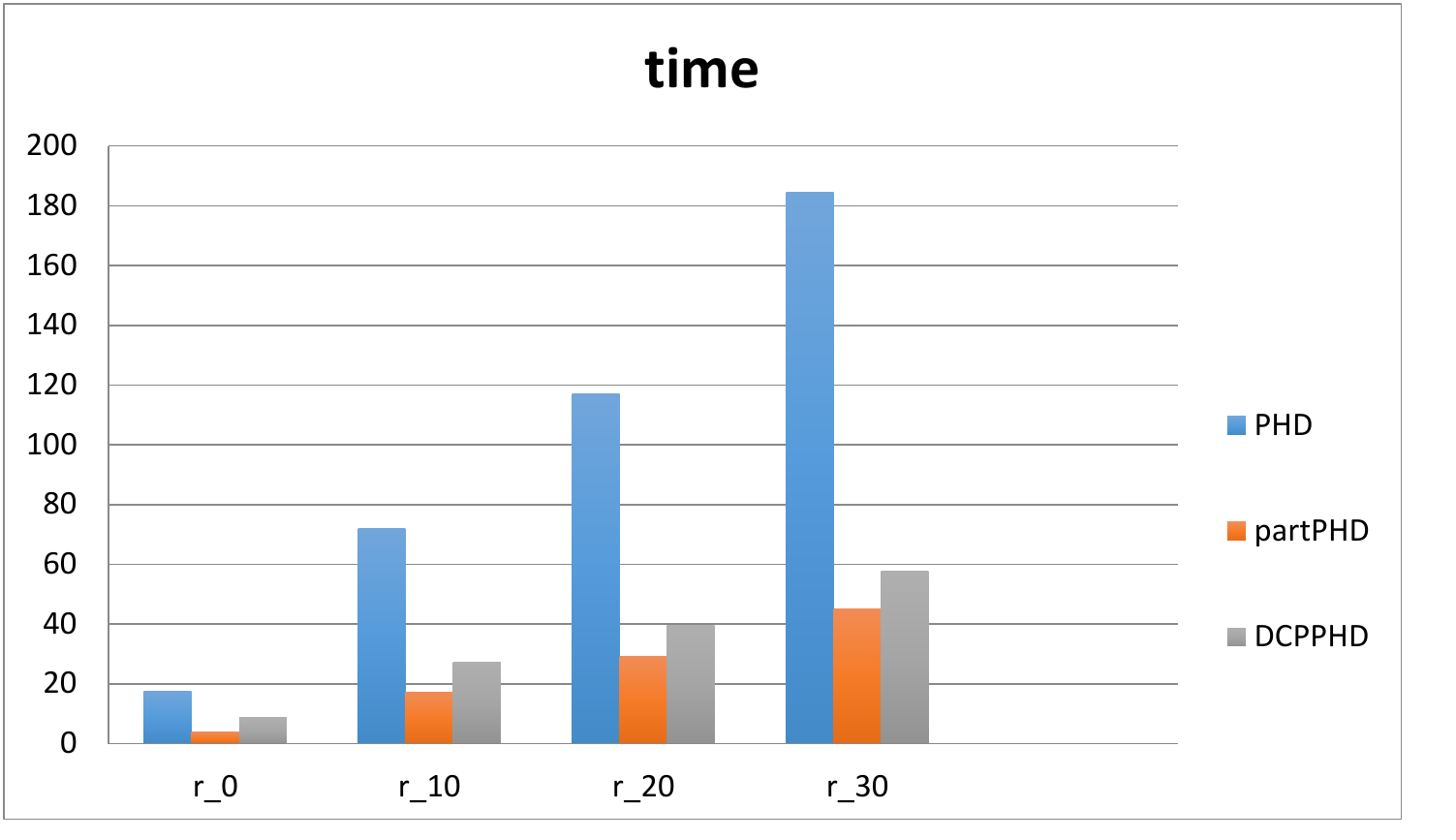}
\caption{the time }
\end{figure}

and analyse the result.

\section{CONCLUSION}In this paper,we proposed a DRNA particle PHD  filter that want to improve the particle PHD filter runtime.The DRNA-PHD filter can speed up the particle PHD filter in theory.However, the feasibility of the proposed method needs to be tested in real applications.It note that divide the more groups will lead to the decrease of performance.\\
future work may consider other particle exchange method and simplify the update step as we found the weight update also cause a lot of compute time.The update step and resamping will be the bottleneck for the development of higher speed of particle PHD filter.And we will test the method into the GPU as well.
\label{}

\bibliographystyle{abbrv}
\bibliography{sigproc}  

\end{document}